\documentstyle[aps,prb,preprint,epsf]{revtex}

\begin{document}
\author{A. G. Rojo}
\title{Electron--drag effects in coupled electron systems }
\address{Department of Physics, The University of
Michigan,
Ann Arbor, Michigan 48109-1120\\
}
\maketitle
\date{The Date }

\begin{abstract}

The advancement of fabrication and lithography techniques of semiconductors have made
it possible to study bi--layer systems made of two electronic layers separated by  distances
of several hundred Angstroms. In this situation the electrons in layer 1 are distinguishable
from those in layer 2, and can communicate through the direct  inter--layer Coulomb interaction.
In particular, if a current is applied to one of the layers, the electrons in the second will
be dragged giving rise to a transresistance $\rho_D$. 
In this article we review recent  theoretical and experimental developments in the understanding
of this effect. At very low temperatures it turns out that phonons dominate the transresistance. 
The direct Coulomb interaction and plasmon excitations are important at temperatures $T>0.1T_F$, with
$T_F$ the Fermi temperature. If a magnetic field is applied the transresistance is increased, in a
very interesting interplay between $\rho_D$ and Landau quantization. The non--dissipative drag 
is also reviewed. 

\end{abstract}

\newpage

\section{introduction}

\bigskip 

\bigskip Electron--electron (e-e)  interactions are responsible for a
multitude of \ fascinating effects in condensed matter. They play a leading
role in phenomena ranging from high temperature superconductivity and the
fractional quantum Hall effect, to Wigner crystalization, the Mott
transition and Coulomb gaps in disordered systems.  The effects of this
interaction on transport properties, however, are difficult to measure.\ \ A
new technique has\ recently proven effective in measuring the scattering
rates due to the Coulomb interaction directly\cite{gramila/prl}. 

This technique is based on an earlier proposal by Pogrebinski\u{\i}\cite
{pogre,price0}. The prediction was that for two conducting systems separated
by an insulator (a semiconductor--insulator--semiconductor layer structure
in particular) there will be a drag of carriers in one film due to the
direct Coulomb interaction with the carriers in the other film. If layer $2$
is an ``open circuit'', and a current starts flowing  in layer $1,$ \ there
will be momentum transfer to layer  $\ 2$ that will start sweeping carriers
to one end of the sample, and inducing a charge imbalance across the film.
The charge will continue to accumulate until the force of the resulting
electric field balances the frictional force of the interlayer scattering.
In the stationary state there will be an induced, or drag voltage $V_{D}$ in
layer $2$. \ 

\bigskip A fundamental difference emerges in the configuration of the
current drag experiment with that in which the in--plane resistance
is measured. For a perfectly pure,
translationally invariant system, the Coulomb interaction cannot give rise
to resistance since the total current commutes with the Hamiltonian $H$. 
This means that states with a finite current are stationary states of $H$
and will never decay, since the e-e interaction conserves not only the total
momentum but also the total current. (For electrons moving in a periodic
lattice, momentum and velocity are no longer proportional and the current
could in principle decay by the e--e interaction.)  If the layers are
coupled by the Coulomb interaction, the stationary states correspond to
linear superposition of states in which the current is shared \ in different
amounts between layers: the total current within a given layer is not
conserved and can relax via the inter--layer interaction.

This mechanism of current degrading was studied in the pioneering experiment
of Gramila {\it et al.}\cite{gramila/prl} for GaAs layers embedded in AlGaAs
heterostructures. \ The separation between the layers was in the range $200$-%
$500$\AA . The coupling of electrons and holes\cite{sivan} and the coupling
between a two dimensional and a three dimensional was also examined\cite
{solomon1}. 

If we call $I$ the current circulating in layer $1$, the drag resistance (or
transresistance) is defined as
\[
\rho _{D}=\frac{V_{D}}{I}.
\]

The naive expectation of the temperature dependece of $\rho _{D}$ is that it
should vanish as $T^{2}$ at low temperatures. This results from the
exclusion principle which limits the scattering to states within $k_{B}T$
of the Fermi surface, and there is one factor of $T$ per layer. \ The first
experiment seemed compatible with the e--e mechanism with some discrepancy
since the observed ratio $\rho _{D}/T^{2}$ was not \ a constant at low
temperatures. This experiment motivated a rather extensive theoretical
effort to understand in better detail the mechanisms of cross--talking
between layers between which electrons are not allowed to tunnel. The theory
in turn stimulated new experiments and in the recent years we have seen
Coulomb Drag developing as a field in itself. \ In the present article we 
briefly review the progress made in understanding different aspects of the
problem. Section \ref{interlayer} provides some background on the Coulomb
interaction and screening for double layer systems. \ In Section \ref
{frictional} we discuss the perturbative equation\ for $\rho _{D}$ used in
most treatments. This equation can be derived in various ways. In the
Appendix we present the memory function derivation and the detailed low
temperature dependence assuming Coulomb scattering only. 

The following sections discuss the crossover between three regimes. Phonon
exchange, discussed in Section \ref{phonons}, dominates the drag at $%
T<0.1T_{F},$ with $T_{F}$ the Fermi temperature. For $T\approx 0.1T_{F\text{ 
}},$ single particle Coulomb scattering is the dominant effect, and for $%
T>0.2T_{F}$, plasmons (See section \ref{plasmons}) are responsible for an
enhancement of the drag current. The theoretical studies of the effects of
disorder and localization are reviewed in Section \ref{disorder}. The
enhancement of $\rho _{D}$ due to an applied magnetic field perpendicular to
the layers and the interplay between Landau quantization and interlayer
scattering is reviewed in Section \ref{magneto}. Finally in Section \ref
{nondisip} we discuss the theory of non-dissipative drag and some
experimental attempts to measure it.

\section{Inter-layer Coulomb interaction}

\label{interlayer}

The  leading actor in this play  is the Coulomb interaction between
electrons in different  layers.  This  interaction is responsible for the
scattering, and will be screened by the density fluctuations within each layer.
If one neglects the interaction between layers, the effective interaction $%
V(q,\omega )$ within a given layer calculated in the Random Phase
Approximation (RPA) is given by \cite{nozieres} 
\begin{equation}
V(q,\omega )=\frac{V_{b}(q)}{1+V_{b}(q)\chi (q,\omega )},  \label{rpa}
\end{equation}
with $V_{b}(q)=2\pi e^{2}/q$ the bare interaction and $\chi (q,\omega )$ the
function characterizing the response of the charge $\delta \rho (q,\omega )$
to an external potential $\varphi (q,\omega )$: $\delta \rho (q,\omega
)=-\chi (q,\omega )\varphi (q,\omega )$. For two coupled layers we also have
the bare interlayer interaction $U_{b}(q)=2\pi e^{2}\exp (-qd)/q$, with $d$
the distance between layers. In addition, the response functions could be
different if the layers are not identical. The effective interaction can be
written now as a $2\times 2$ matrix\cite{santoro,dassarma,ussishkin} 
\begin{equation}
\hat{V}(q,\omega )=\hat{V}_{B}(q)\frac{1}{1+\hat{\chi}(q,\omega )\hat{V}%
_{B}(q)}
\label{vqomega}
\end{equation}
with $[\hat{V}_{B}(q)]_{11}=[\hat{V}_{B}(q)]_{22}=V_{b}(q),$ $[\hat{V}%
_{B}(q)]_{12}=[\hat{V}_{B}(q)]_{21}=U_{b}(q)$, and $[\hat{\chi}(q,\omega
)]_{ij}=\delta _{ij}\chi _{i}(q,\omega )$, with $i=1,2$ labeling the layer.
Note that $\chi_1$ and $\chi_2$ could be different, so Eq. (\ref{vqomega})
is valid for non-identical layers.
The off--diagonal element of $\hat{V}(q,\omega )$ is the screened interlayer
interaction

\begin{equation}
\lbrack \hat{V}(q,\omega )]_{12}=\frac{U_{b}(q)}{\left[ 1+\chi _{1}(q,\omega
)V_{b}(q)\right] \left[ 1+\chi _{2}(q,\omega )V_{b}(q)\right]
-U_{b}^{2}(q)\chi _{1}(q,\omega )\chi _{2}(q,\omega )}\equiv
U^{(S)}(q,\omega )  \label{effint}
\end{equation}

The zeroes of the denominator in the above equation correspond to the
collective modes of \ charge \ oscillations. It is illustrative to write the
screened interaction for identical layers in the form (omitting the argument 
$q$ in the interactions for simplicity) 
\begin{equation}
U^{(S)}(q,\omega )=\frac{1}{2}\left[ \frac{V_{b}+U_{b}}{1+\chi (q,\omega
)(V_{b}+U_{b})}-\frac{V_{b}-U_{b}}{1+\chi (q,\omega )(V_{b}-U_{b})}\right] 
\label{poles}
\end{equation}
which puts in evidence the symmetric and antisymmetric modes. The response
function $\chi (q,\omega )$ has real and imaginany parts. The imaginary part
corresponds to the continuum of particle--hole excitations that exist for
frequencies $\omega <v_{F}q$. The collective modes corresponding to the
poles of $U^{(S)}(q,\omega )$ will \ be non--decaying at $T=0$ if they fall
out of the particle--hole continuum, which guarantees that the poles are in
the real axis. We can find these stable modes by expanding $\ $the real part
of $\ \chi (q,\omega )$ \ for the free electron gas for small wave vectors
and $\omega \gg v_{F}q$: 
\begin{equation}
\chi (q,\omega )\simeq -\frac{n}{m}\left( \frac{q}{\omega }\right) ^{2}
\label{smallq}
\end{equation}
with $m$ the electron mass, and $n$ the electron density\cite{stern}. In the small $q$
limit, $V_{b}+U_{b}\approx 4\pi e^{2}/q$, and $V_{b}-U_{b}\approx 2\pi
e^{2}d.$ If we substitute this in Eq. (\ref{poles}) we find that the two
poles are at frequencies $\omega _{\pm }(q)$ given by 
\begin{eqnarray*}
\omega _{+}(q) &=&e\sqrt{\frac{4\pi n}{m}}\sqrt{q}, \\
\omega _{-}(q) &=&e\sqrt{\frac{2\pi nd}{m}}q.
\end{eqnarray*}

This in--phase and out--of--phase modes are usually labeled ``optical
plasmon'' and ``acoustic plasmon'' respectively, and are stable at $T=0$. At
finite temperatures $\chi (q,\omega )$ acquires a finite imaginary part
outside of the particle--hole continuum, the poles move into the complex
plane and the plasmons acquire a finite lifetime. In section \ref{plasmons}
we will discuss the role of plasmons in enhancing the drag resistance at
temperatures of the order of the Fermi energy.

\section{Frictional drag}

\label{frictional}

The cross resistance is calculated in most of the recent theoretical
treatments from the following expression, valid when the interlayer
interaction is treated perturbately\cite{sivan}(See Figure 1):

\begin{equation}
\rho _{D}=-{\frac{\hbar ^{2}\beta }{\pi n_{1}n_{2}e^{2}A}}\int {\frac{dq}{%
(2\pi )}}q^{3}|U^{(S)}(q,\omega )|^{2}\int_{-\infty }^{\infty }d\omega {%
\frac{{\rm 
\mathop{\rm Im}%
}\chi _{1}(q,\omega ){\rm 
\mathop{\rm Im}%
}\chi _{2}(q,\omega )}{\sinh ^{2}(\beta \hbar \omega /2)}}.  \label{central}
\end{equation}

In Appendix A we reproduce the derivation by of this equation by Zheng and
MacDonald who used the memory function formalism \cite{zheng}. 
This derivation is  general and contains Equation (\ref{central}) as the 
limit of lowest order in the inter--layer interaction. 
We will show here 
a more restricted derivation using the scattering formalism, which has the
advantage of providing us with an interpretation of the origin of each
of the terms in Equation (\ref{central}).
The derivation is very close to the one presented in the  
 paper by Jauho and Smith\cite{jauho,solomon}. 
 Alternative treatments are the collective excitation approach\cite
{maslov}, and the Green's function formalism staring from the Kubo formula
\cite{tso,hansch,flensberg0}. In the linearized Boltzman (or scattering) 
treatment, the
spectral functions $\chi _{i}(q,\omega )$ are the susceptibilities of the
non--interacting case.

\vspace {2.cm}
We compute  first the rate of change of  the momentum of the electrons in
layer 2 due to the scattering from electrons in layer 1:
\begin{eqnarray}
{d{\rm \bf P}_2\over dt}&=& \int {d{\bf q}\over (2\pi)^2}
\hbar {\bf q}|U(q)|^2 \int {d{\bf k}_1\over (2\pi)^2}\int {d{\bf k}_2\over (2\pi)^2}
\left\{ f_{{\bf k}_1}(1-f_{{\bf k}_1+{\bf q}}) 
 f_{{\bf k}_2}^0(1-f_{{\bf k}_2-{\bf q}}^0) \right.
 \nonumber
\\
& &-\left. 
f_{{\bf k}_1+{\bf q}}(1-f_{{\bf k}_1})
 f_{{\bf k}_2-{\bf q}}^0(1-f_{{\bf k}_2}^0)
\right\}\delta (\epsilon_{{\bf k}_1}+\epsilon_{{\bf k}_2}
-
\epsilon_{{\bf k}_1+{\bf q}}-\epsilon_{{\bf k}_2-{\bf q}}).
\label{dpdt}
\end{eqnarray}

In the above equation, $f_{\bf k}$ refers to the distribution of electron states in layer 1,
and $f_{\bf k}^0$ to that of the electrons in layer 2. Since there is no current in layer 2,
the corresponding 
electron distribution is the unperturbed, free--fermion equilibrium distribution.  
The electron distribution on layer 1, on the other hand, corresponds to a  
 Fermi distribution displaced in $k$--space by an amount $m{\bf v}_1/\hbar$, with 
 ${\bf v}_1$ being the drift velocity of the electrons in layer 1. 
 The above expression can be interpreted as a sum of procesess according to  Fermi's golden rule:
 the factor  $|U(q)|^2$ is the square of the matrix element of the transition that involves
 a momentum transfer $\hbar {\bf q}$, the various factors of $f$ correspond to the probabilities
 of transitions from occupied to empty states, and the delta function ensures conservation of
 energy in the scattering process. Note that if layer 1 were in equilibrium, the magnitude in
 curly brackets is zero due to the detailed balance condition
 \begin{equation}
 f_{{\bf k}}^0
 (1-f_{{\bf k}'}^0)
 f_{{\bf k}''}^0
 (1-f_{{\bf k}'''}^0)
 =
 f_{{\bf k}'}^0
  (1-f_{{\bf k}}^0)
   f_{{\bf k}'''}^0
    (1-f_{{\bf k}''}^0),
 \end{equation}
 which is an identity if $\epsilon_{{\bf k}}+\epsilon_{{\bf k}'} +\epsilon_{{\bf k}''}+ \epsilon_{{\bf k}'''}=0$.
 There is therefore a finite momentum transfer to layer 2 due to the asymmetry of  the electron distribution
 in layer 1.

We now linearize in the electron distribution of layer 1:
\begin{equation}
f_{\bf k} =f_{{\bf k}-m {\bf v}_1/\hbar}^0= f_{\bf k}^0 - {\partial f_{\bf k}^0
\over \partial \epsilon_{\bf k}} 
\hbar {\bf k}\cdot {\bf v}_1\equiv f_{\bf k}^0
-{1\over k_B T} f_{\bf k}^0(1-f_{\bf k}^0)\hbar {\bf k}\cdot {\bf v}_1.
\end{equation}

Subtituting the linearized expressions for  $f_1$ in Equation (\ref{dpdt}),
and using the detailed balance condition the rate of momentum transfer 
has the form
\begin{equation}
{d{\rm \bf P}\over dt}={{\bf v}\over  2} {1\over k_B T}
\int {d{\bf q}\over (2\pi)^2}(\hbar  q)^2|V(q)|^2 \int {d{\bf k}_1\over (2\pi)^2}\int {d{\bf k}_2\over (2\pi)^2}
f_{{\bf k}_1}^0(1-f_{{\bf k}_1+{\bf q}}^0)
 f_{{\bf k}_2}^0(1-f_{{\bf k}_2-{\bf q}}^0)\delta (\epsilon_{{\bf k}_1}+\epsilon_{{\bf k}_2}
 -
 \epsilon_{{\bf k}_1+{\bf q}}-\epsilon_{{\bf k}_2-{\bf q}}).
 \label{dpdt2}
\end{equation}

A couple of simple technical manipulations are needed to recast this expression in the form
of our basic equation. First we use 
\begin{equation}
\delta (\epsilon_{{\bf k}_1}+\epsilon_{{\bf k}_2}
 -
  \epsilon_{{\bf k}_1+{\bf q}}-\epsilon_{{\bf k}_2-{\bf q}})
  =\int \hbar d\omega \delta (\hbar \omega -\epsilon_{{\bf k}_1}+\epsilon_{{\bf k}_1+{\bf q}})
  \delta ( \hbar \omega + \epsilon_{{\bf k}_2} -\epsilon_{{\bf k}_2-{\bf q}}),
\end{equation}
and
\begin{equation}
f^0(\epsilon_{{\bf k}})\left[1-f^0(\epsilon_{{\bf k}}+\hbar \omega)\right]= 
\left[f^0(\epsilon_{{\bf k}})-f^0(\epsilon_{{\bf k}}+\hbar \omega)
\right]/\left[1-\exp(-\hbar \omega/k_BT)\right].
\end{equation}
We then introduce the function $\chi^0_i(q,\omega)$:
\begin{equation}
 \int {d{\bf k}_i\over (2\pi)^2} (f_{{\bf k}_i}^0-f_{{\bf k}_i+{\bf q}}^0) 
\delta (\hbar \omega -\epsilon_{{\bf k}_1}+\epsilon_{{\bf k}_1+{\bf q}})
={1\over \pi}{\rm Im} \int
{d{\bf k}_i\over (2\pi)^2} {f_{{\bf k}_i}^0-f_{{\bf k}_i+{\bf q}}^0
\over \hbar \omega -\epsilon_{{\bf k}_1}+\epsilon_{{\bf k}_1+{\bf q}}-i\eta}
\equiv {\rm Im} \chi_i^0(q,\omega),
\end{equation}
and  substitute in (\ref{dpdt2}):
\begin{equation}
{d{\rm  P}_2\over dt}={{ v}_1\over  2} {1\over k_B T}
\int {d{\bf q}\over (2\pi)^2}(\hbar  q)^2|V(q)|^2
\int d\omega
{{\rm Im}\chi^0_1(q,\omega) 
{\rm Im}\chi^0_2(q,\omega) \over \sinh ^2(\hbar\omega/k_BT)},
\end{equation}
where we have omitted the vectorial nature of ${\rm \bf P}_2$ and ${\bf v}_2$ since
they are in the same direction.
In order to compute the resistance, we need to equate the rate of momentum transfer
to the total force  per particle on the electrons in layer 2 due to the generated
electric field $E_2$:
 \begin{equation}
 {d{\rm  P}_2\over dt}= + n_2 e E_2,
 \end{equation}
 and we get the final expression using $\rho _D=E_2/j_1$ with $j_1=n_1 e v_1$.

The minus sign in Eq. (\ref{central})  means that the induced
drag voltage is opposite to the resistive voltage drop in the current
carrying layer. This is so because the Coulomb-induced scattering sweeps the
carriers along the dragged layer in the same direction as those in the drive
layer. From the above equation we see that the drag resistance is a
convolution of the density fluctuations within each layer, which at low
temperatures are restricted to low frequencies by the factor $\sinh
^{2}(\beta \hbar \omega /2)$ in the denominator. From the structure of the
above equation we see that measurements of $\rho _{D}$ can provide
information on the inter--layer scattering mechanism as well as the
in--plane fluctuations, the information on which is included in $\chi
(q,\omega )$. The low temperature prediction of Coulomb scattering alone is
\cite{gramila/prl} 
\begin{equation}
\rho _{12}={\frac{m}{ne^{2}}}{\frac{\pi \zeta (3)(k_{B}T)^{2}}{16\hbar
E_{F}(q_{{\rm TF}}d)^{2}(k_{F}d)^{2}},}
\end{equation}
where $q_{{\rm TF}}$ is the single-layer Thomas-Fermi screening wavevector, $%
E_{F}$ is the Fermi energy, $d$ is the interlayer spacing and $k_{F}$ is the
Fermi wave vector. Due to the finite separation of the layers, the
scattering is limited to small angles\cite{gramila0}. Large angle scattering
events require large values of the momentum transfer $q$, and these
processes are suppressed by the exponential dependence $\sim e^{-qd}$ in the
Fourier transform of the inter--layer Coulomb interaction. In two
dimensions, the phase space for backscattering diverges and gives rise to
logarithmic corrections to the simple $T^{2}$ dependence\cite{hodges}. This
corrections are missing here due to the suppresion of backscattering. The
screening is also enhanced at small wave vectors, and is more effective as
the separation between layers increases [see Eq (\ref{screened})]. The
combination of the suppression of backscattering and the enhanced screening
gives rise to the strong $d^{-4}$ dependence in $\rho _{D}$. 
It shoud be emphasized also that the $T^^2$ dependence is modified when the electronic states in the
layers are not free Fermi gasses, and support longer-lived density
fluctuations. As detailed in the following sections, this is the case,
{\it e.g.}, in the presence of disorder, and in the quantum Hall regime."

\section{\protect\bigskip The role of phonons}

\label{phonons} \ 

The electron--electron scattering treatment predicts a transresistance $\rho
_{D}$ that vanishes at zero temperature as $\rho _{D}\sim T^{2}$. This
dependence is roughly satisfied in the experiments, confirming the dominance
of the electron--electron interaction. However, as mentioned above, the
experiment also shows a noticeable deviation of $\rho _{D}/T^{2}$ from a
constant as a function of temperature, showing a maximum for $T\sim 2K.$
For this samples the Fermi temperature is roughly $ 60 K$. The
overall temperature dependence and the position of the maximum are very
similar for different values of the layer separations $d$, and the magnitude
of $\rho _{D}/T^{2}$ varies very little for three barrier thickness $%
d=500,225$ and $500$\AA\ after the Coulomb scattering contribution was
subtracted\cite{gramila0}. Also, for interlayer separations $d=500$\AA , the
observed $\rho _{D}$ is simply too big to be accounted for by the Coulomb
interaction alone. This led Gramila {\it et al}.\cite{gramila0} to propose
an additional scattering mechanism. The obvious candidate:\ phonons. The
phonon--mediated coupling between electrons in doped semiconductor layers
separated by an insulating region of \ thickness $\sim 100\mu $m was studied
by Huber and Shockley\cite{hubner}. Real phonons were found to be
responsible for the interlayer interaction. The possibility of drag due to
phonons was also proposed\ by Gurzhi and Kopliovich\cite{gurzhi}. The first
qualitative hint of the mechanism being phononic is the fact that the
measured temperature dependence of $\rho _{D}/T^{2}$ resembles the
acoustic--phonon--limited mobility scattering rate $\tau _{{\rm ph}}^{-1}$
for two--dimensional electrons in GaAs. \ \ At high temperatures $\tau _{%
{\rm ph}}^{-1}$ is linear in $T$, but crosses over to $T^{5}$ or $T^{7}$ in
the low-temperature Bloch Gr\"{u}niesen regime\cite{price}, where the
thermal phonon wave vectors are less than $2k_{F}.$ \ For the electron
density \ of the samples in Ref. \cite{gramila0} the crossover occurs at a
few degrees Kelvin. Thus, the temperature dependences of $\tau _{{\rm ph}%
}^{-1}/T^{2}$ and $\rho _{D}/T^{2}$ are broadly similar. Furthermore, the
dependence of $\rho _{D}/T^{2}$ on the relative electron density between the
layers indicates that phonons could be playing a role. An electron in layer
1 decays through a backscattering process by emitting a phonon of
wave--vector $2k_{F,1}$, and the phonon will transfer its momentum most
efficiently to an electron in layer 2 if $k_{F,1}=k_{F,2}$. This implies
that $\rho _{D}/T^{2}$ should have a maximum when the electron densities are
matched in the two layers, which is experimentally observed\cite{gramila0}.
Now, interactions of acoustic phonons with electrons are relatively weak in
GaAs to account for the value of the observed transresistance. The proposed
mechanism will be the exchange of virtual phonons\cite{zhang1,tso}, a
process in which a phonon is emitted by one layer and then absorbed by the
second without conserving energy from the electronic transitions. When the
energy conservation constraint is relaxed, the phase space for scattering
increases. Also, since the layers are separated by distances much smaller
than the phonon mean free path, the phonons retain their phase coherence for
the interaction between the layers. This two effects imply an enhancement in
the transresistance due to virtual phonons. Tso {\it et al.}\cite{tso0}
presented diagrammatic calculations including exchange of virtual phonons
with a good agreement with the temperature dependence observed in the
experiment (See Figure 2).

\bigskip The distinction between ``real'' and ``virtual''\ phonons is not so
clear cut if one extends the treatment leading to Eq. (\ref{central}) to
include a phonon--mediated Coulomb interaction. The force operator on a
given layer will involve the phonon operators, and the force--force
correlation function will include an electron--phonon interaction $M(Q)$ and
a the propagator of a phonon from layer one to layer two. The result is that
one can write Eq. (\ref{central}) with $U^{(S)}(q,\omega )$ replaced by an
interaction ${\cal D}_{12}(q,\omega )$ of the form\cite{bonsager3} 
\begin{equation}
{\cal D}_{12}(q,\omega )=\int \frac{dq_{z}}{2\pi \hbar }|M(Q)|^{2}e^{iQ_{z}d}%
\left[ \frac{2\omega _{Q}}{\omega ^{2}-\omega _{Q}^{2}}\right] ,
\label{phononprop}
\end{equation}
with $\omega _{Q}=v_{s}Q$ the frequency of the acoustic phonon (omitting the
distinction between longitudinal and transverse) of wave vector $Q=\sqrt{%
q^{2}+Q_{z}^{2}}.$ The term in square brackets in the above equation is the
phonon Green's function\cite{mahan}. The phonon mean free path $\ell _{ph%
\text{ }}$is not included. In the above expression, one can interpret $\hbar
\omega $ as the energy transferred between layers and $\hbar \omega _{Q}$ as
the energy of the intermediate phonon. When the denominator in the square
bracket of Eq. (\ref{phononprop}) vanishes, energy is conserved in the
intermediate state. As pointed out by B\o nsager {\it et al.}\cite{bonsager3}%
, this expression contains both the real and virtual phonons: real phonons
correspond to $%
\mathop{\rm Im}%
{\cal D}_{12}$, whereas virtual phonons correspond to $%
\mathop{\rm Re}%
{\cal D}_{12}.$ If we insert in Eq. (\ref{phononprop}) the long wave-length
limit\cite{price2} $|M(Q)|^{2}=\hbar QD^{2}/2\rho v_{s}$, we obtain\cite
{bonsager3} 
\begin{equation}
{\cal D}_{12}(q,\omega )\approx \frac{\omega ^{2}}{q\sqrt{1-\omega
^{2}/(v_{s}q)^{2}}}\exp \left( -qd\sqrt{1-\omega ^{2}/(v_{s}q)^{2}}\right) ,
\label{divergente}
\end{equation}
indicating that the effective interaction diverges as $q\rightarrow \omega
/v_{s}.$ Even though ${\cal D}_{12}(q,\omega )$ involves a small prefactor,
if we substitute in Eq.(\ref{central}) the obtained expression for the
coupling, the divergence $|{\cal D}_{12}(q,\omega )|^{2}\sim |\omega
-v_{s}q|^{-1}$ gives rise to a divergent $\ \rho _{D}$. Although this is a
spurious divergence that is removed by inclusion of a finite mean free path
of the phonons or dynamical screening of the interaction, the large
contribution explains partly why a ``weak'' phonon mediated interaction can
compete with the Coulomb interaction as a mechanism for drag. For a phonon
mean free path $\ell _{{\rm ph}}$ below a critical value which for GaAs can
be of the order of $0.2$mm, the predicted distance dependence of the drag is 
$\ln (\ell _{{\rm ph}}/d).$ The experimental distance dependence is yet to
be clarified, although the evidence is to a weak dependence of the phonon
contribution\cite{lilly}.

\section{The effect of disorder and localization}

\label{disorder}

The effect of disorder was first studied by Zheng and MacDonald\cite{zheng},
who included the fact that the density response function at small
frequencies and small wave-vectors is given by 
\begin{equation}
\chi _{i}\left( q,\omega \right) =\frac{dn}{d\mu }\frac{Dq^{2}}{%
Dq^{2}-i\omega },  \label{diffusive}
\end{equation}
with $D$ being the diffusion constant\cite{leerama} given by $D=l^{2}/2\tau $%
, $l$ being the mean free path, $\tau =l/v_{F\text{ }}$ the scattering time,
and $dn/d\mu $ the density of states . This formula for the density response
function is valid for $q<1/l$ and $\omega <1/\tau $. Following the
derivation of $\rho _{D}\sim T^{2}$ presented in Appendix A, Zheng and
MacDonald obtain that the low temperature dependence is modified to $\rho
_{D}\sim T^{2}\log T$. This temperature dependence can be seen to
result from the low frequency and low wave--vector behavior of $%
\mathop{\rm Im}%
\chi _{i}\left( q,\omega \right) \sim \omega /q^{2}$ for $q>(\omega
/D)^{1/2}.$ (Note that in the ballistic regime $%
\mathop{\rm Im}%
\chi _{i}\left( q,\omega \right) \sim \omega /q.$) The contributions to the
integral in $q$ for $q<(\omega /D)^{1/2}$ can therefore be neglected. We can
obtain the low temperature behavior of $\rho _{D}$ as follows\cite
{zheng,kamenev}: 
\begin{equation}
\rho _{D}\sim \frac{e^{2}}{(k_{{\rm TF}}d)^{2}}\beta \int_{0}^{1/\tau
}d\omega \frac{\omega ^{2}}{e^{\beta \hbar \omega }+e^{-\beta \hbar \omega
}-2}\int_{(\omega /D)^{1/2}}^{1/l}\frac{dq}{q}\approx \frac{e^{2}}{(k_{{\rm %
TF}}d)^{2}}T^{2}\log T.
\end{equation}

In the above expression we have considered the low wave--vector contribution
from the Coulomb potential as a constant $U^{(S)}(q,\omega )\sim e^{2}/[k_{%
{\rm TF}}(k_{{\rm TF}}d)^{2}]$. Note the change in distance dependence of $%
\rho _{D}$ from $d^{-4}$ to $d^{-2}$. This is a consequence of having
treated the Coulomb scattering as a screened interaction.
The logarithmic term originates in the different spectrum of 
particle--hole excitations for a ballistic from a diffusive system
and from the dimensionality of the system. 
Both the spectrum of excitations and the dimensions (two in this case)
conspire to give the logatrithmic term in the temperature dependence.   
At low
temperatures one can ignore the contribution of $%
\mathop{\rm Im}%
\chi $ to the screened potential given by Equation (\ref{effint}), and
replace the interlayer interaction by 
\begin{equation}
U^{(S)}(q,\omega )=\frac{\pi e^{2}q}{k_{{\rm TF}}^{2}\sinh qd},
\label{screened}
\end{equation}
with $k_{{\rm TF}}\equiv 2\pi e^{2}dn/d\mu $ the single layer Thomas--Fermi
screening wave vector.

The case of strong disorder, where the localization length $\xi $ the of
states within each layer is of the order of the distance between layers was
considered by Shimshoni\cite{shimshoni1}. For the case of Anderson
insulators, the density response function within each layer is taken from
the self consistent theory of localization of Wollhardt and W\"{o}lfle\cite
{wollhardt}: 
\begin{equation}
\chi (q,\omega )=\frac{dn}{d\mu }\frac{Dq^{2}}{Dq^{2}-[i\omega +\tau (\omega
^{2}-\omega _{0}^{2})]},
\end{equation}
with $\omega _{o}$ a restoring frequency that incorporates the effects of
localization. Within the scheme of Wollhardt and W\"{o}lfle, $\omega _{0}$
is related to the localization length $\xi $ through $\xi =v_{F}/(\sqrt{2}%
\omega _{0}).$ The parameter $\omega _{0}$ in the above expression is
responsible for the vanishing quadratically of the conductivity $\sigma (\omega )$
 at low frequencies: 
\begin{equation}
\mathop{\rm Re}%
\sigma (\omega )=e^{2}\lim_{q\rightarrow 0}(\omega /q^{2})%
\mathop{\rm Im}%
\chi (q,\omega )\sim \lbrack D/(\tau ^{2}\omega _{0}^{4})]\omega ^{2}.
\end{equation}

Note that the logaritmic divergence at low $q$ is now cut-off by $1/\xi $
and the $T^{2}$ dependence is retained. In the strongly localized regime,
when the localization length is of the order of the inter--layer spacing,
the distance dependence of $\rho _{D}$ is modified. Screening is not
effective in this regime, so we can evaluate $\rho _{D}$ using the
unscreened interaction $U_{{\bf q}}=2\pi e^{2}e^{-qd}/q.$ In this regime,
the low $q$ dependence $%
\mathop{\rm Im}%
\chi (q,\omega )$ will be given by 
\begin{equation}
\mathop{\rm Im}%
\chi (q,\omega )\approx \frac{dn}{d\mu }\frac{\omega q^{2}}{D[1/\xi
^{2}+q^{2}]^{2}}\approx \frac{dn}{d\mu }\frac{1}{D}\omega q^{2}\xi ^{4}.
\end{equation}

Replacing the above expression in Equation (\ref{central}) we obtain\cite
{shimshoni1}, 
\begin{equation}
\rho _{D}\approx \frac{5}{32\pi }\frac{h}{e^{2}}%
{k_{B}T \overwithdelims() \hbar Dn}%
^{2}(k_{{\rm TF}}d)^{2}%
{\xi  \overwithdelims() d}%
^{8}.
\end{equation}

If the localization length $\xi \gg d$ one has to include the effects of
screening for wave--vectors $q>1/(\xi ^{2}k_{TF})$, and the distance
dependence changes to $\ln (k_{{\rm TF}}$ $\xi )/(k_{{\rm TF}}d)^{2}.$ When $%
\xi \rightarrow \infty $ we recover the diffusive result. Another case
treated in Ref (\cite{shimshoni1}) is that of the so called
Efros--Shklovskii insulators, where Coulomb interactions are important and
increase the conductivity at finite frequency in such a way that $\sigma
(\omega \rightarrow 0)\sim \omega $ as opposed to $\omega ^{2}.$ This has
the consequence that the integration in $\omega $ is infrared divergent and
is cut-off, at finite $T$ by incoherent phonon processes. As a consequence
the drag resistance diverges as zero temperature in the following way 
\[
\rho _{D}\propto f(d,\xi )%
{T \overwithdelims() T_{0}}%
^{3}\exp \left[ 
{T_{0} \overwithdelims() T}%
^{1/2}\right] ,
\]
with $f(d,\xi )=1/(\xi d)^{2}$ for $\xi \ll d$ and $f(d,\xi )=1/\xi ^{4}$
for $\xi \gg d$. Also $k_{B}T_{0}=e^{2}/\epsilon \xi $, with $\epsilon $ the
dielectric constant.
\newpage
\section{\protect\bigskip\ Plasmon enhancement}

\label{plasmons}

The discussion of plasmon modes of Section \ref{interlayer} focuses on the
small $\ q$ dispersion at zero temperature. Since the drag resistance is
given by an integral involving $%
\mathop{\rm Im}%
\chi $ in the integrand, at low temperature there is no contribution from
the plasmons, since they appear in a region of the plane $(q,\omega )$ where 
$
{\rm Im} \chi= 0$. 
 The function ${\rm Im}\chi $  ``counts" the number of particle
 hole excitations of mementum $a$ and frequency $\omega$: it 
 is nonzero in regions of the $(q,\omega)$ plane where particle--hole excitations
 are allowed. The plasmons are collective modes that are outside of the 
 particle--hole continuum. For a more detailed discussion of this point the 
 reader is referred to Nozi\`eres and Pines\cite{pines}.

\ Flensberg and Hu\cite{flensberg1,flensberg2} made the very
interesting observation that at higher temperatures ($T\sim T_{F})$, $%
\mathop{\rm Im}%
\chi \neq 0$ in the region of the plasmon pole, and therefore there will be
a large contribution at intermediate values of $q$ and one should observe an
enhancement of the resistance\cite{tanatar3}. Qualitatively one can
understand this effect as a Schottky--like peak that develops from a
thermally populated dissipation channel that is not available at zero
temperature. In the plasmon--pole approximation\cite{flensberg2} one
approximates the dielectric constant $\epsilon (q,\omega )$ [which is the
denominator in Eq. (\ref{effint})] by 
\begin{equation}
\epsilon (q,\omega )=2V_{b}(q)e^{-qd}|\beta _{\pm }(q)[\omega -\omega _{\pm
}(q)]+i%
\mathop{\rm Im}%
\chi (q,\omega _{\pm })|,  \label{dielectric}
\end{equation}
with $\beta _{\pm }=[d%
\mathop{\rm Re}%
\chi (q,\omega )/d\omega ]|_{\omega =\omega _{\pm }}.$ For small values of
the imaginary part of $\chi $ one can approximate the Lorenzian by $\delta $
functions and obtain 
\begin{equation}
|U^{(S)}(q,\omega )|^{2}\approx \frac{\pi }{4%
\mathop{\rm Im}%
\chi (q,\omega _{\pm })|\beta _{\pm }(q)|}\delta (\omega -\omega _{\pm }(q)),
\label{plasmon1}
\end{equation}
which we can replace in Eq. (\ref{central}) to obtain the plasmon
contributions to the drag rate: 
\begin{equation}
\rho _{D}={\frac{\hbar ^{2}\beta }{\pi n_{1}n_{2}e^{2}A}}\int_{0}^{q_{c,\pm
}}{\frac{dq}{(2\pi )}}q^{3}{\frac{{\rm 
\mathop{\rm Im}%
}\chi \lbrack q,\omega _{\pm }(q)]}{4|\beta _{\pm }(q)|\sinh ^{2}[\beta
\hbar \omega _{\pm }(q)/2]}}.  \label{plasmondrag}
\end{equation}

The parameter $q_{\pm ,c}$ defines the values of \ $q$ for which the plasmon
ceases to exist. The contribution to the drag from the above equation
involves an integral of $%
\mathop{\rm Im}%
\chi $ for frequencies and wave vectors corresponding to the plasmon modes.
From the above discussion, the contribution of the plasmons to $\rho _{D}$
is zero at $T=0.$ At small temperatures, $%
\mathop{\rm Im}%
\chi $ at $\omega _{\pm }(q)$ is small because the carriers do not have
sufficient energy to be excited \ far above the Fermi surface. However, at
intermediate temperatures (of the order of the Fermi temperature)\ there are
enough thermally excited particles to give a large $%
\mathop{\rm Im}%
\chi $ at the plasmon poles and the drag will be dominated by this
contribution. The numerical calculations indicate a maximum in $\rho _{D}$
at a temperature of the order of $0.5T_{F}.$ For electron densities of the
order of $n=1.5\times 10^{11}{\rm cm}^{-2}$ for GaAs quantum wells, $%
T_{F}\sim 50K$. \ For $T>$ 0.5 $T_{F\text{ }}$the plasmon modes are no
longer well defined since they can decay by emitting single particle
excitations. This implies that the enhancement diminishes at high
temperatures.

\ The plasmon enhancement theory was put to test in the experiments by Hill%
{\it \ et al.}\cite{hill2} with very good qualitative agreement. The
temperature required to excite a plasmon appears to be lower than the value
predicted by the theory, and the magnitude of the drag is larger than the
prediction on most of the temperature range. This seems to imply that one
should go beyond the Random Phase Approximation. The Hubbard approximation 
\cite{flensberg2} provides a better fit to the data, but there is still room
for improvement (see Figure 3).

Improvements over RPA were considered by \'{S}wierkowski {\it et al.}\cite
{swierkowski} in connection with drag experiments in electron--hole systems
by Sivan {\it et al.}\cite{sivan}. The transresistance was calculated using
Eq. (\ref{central}) modifying the effective inter--layer interaction using
the local field approach of Singwi, Tosi, Land and Sj\"{o}lander (STLS)\cite
{singwi,zheng2,swierkowski2}. This approach includes correlations
neglected in RPA,\ and the net effect is an increase of the effective
interaction. This corrections use zero temperature local field corrections.
Tso{\it \ et al.}\cite{tso1} studied a generalized RPA including exchange
processes to infinite order in the Hartree--Fock potential and emphasized
the fact that RPA is good only for very high densities. Thus, the RPA
treatment of the Coulomb scattering overestimates the screening and renders
the coupling weaker. The agreement with the experiments of Ref. \cite{sivan}
is good at low temperatures. G\"{u}ven and Tanatar\cite{guven} treated the
plasmons and the phonons on equal footing using a the Hubbard approximation
and found an enhancement due to the coupled plasmon--phonon modes. We still
need better approximations beyond RPA in the higher temperature regime to
have a better quantitative understanding of the plasmon enhancement effects.

\section{Drag in the presence of magnetic fields}

\label{magneto}

\subsection{Hall drag}

In the presence of a magnetic field applied perpendicular to the planes,
there could in principle exist a trans--Hall resistance, that is, a voltage
in the $y$ direction of layer $2$ when a current is applied in the $x$
direction of layer $1$. This drag Hall resistance is zero if computed it in
the lowest order in the inter--layer interaction.
The reason in that in lowest order the  electron distribution in 
layer 2 is the equilibrium one, and also the fact that the momentum
is being transferred from layer 1 in the direction of the current. 
Since there is no net current in layer 2 there is no Lorentz force and
therefore no net Hall voltage. To see how this  emerges formally
so we can repeat the steps
that lead to Equation \ref{central}. In that case we obtain 
\begin{equation}
\rho _{D}^{xy}={\frac{\hbar ^{2}\beta }{\pi n_{1}n_{2}e^{2}A}}\int {\frac{d%
{\bf q}}{(2\pi )^{2}}}q_{x}q_{y}|U^{(S)}(q,\omega )|^{2}\int_{-\infty
}^{\infty }d\omega {\frac{{\rm 
\mathop{\rm Im}%
}\chi _{1}({\bf q},\omega ){\rm 
\mathop{\rm Im}%
}\chi _{2}({\bf q},\omega )}{\sinh ^{2}(\beta \hbar \omega /2)}},
\label{drag}
\end{equation}
where the functions $\chi _{1}({\bf q},\omega )$ have to be evaluated in the
presence of a magnetic field ${\bf B(x)=\nabla \times A(x}_{i}).$ The reason
why the equation retains its form is the fact that the force operator is
still given by Equation (\ref{force}) even in the presence of a magnetic
field since $[{\bf A(x}_{i}),\rho _{{\bf q}}]=0$. In other words, the
interlayer force does not depend on the applied field. Due to rotational
invariance $\chi _{1}({\bf q},\omega )=\chi _{1}(|{\bf q|},\omega )$ and the
integral above vanishes by parity. The result $\rho _{D}^{xy}=0$ was also
shown by Kamenev and Oreg\cite{kamenev} using a diagrammatic approach.
However, as emphasized by Hu\cite{byk} this is not a general result. In
particular, Hu showed that using an approach based on the Boltzman equation,
if one includes an {\em energy dependent} lifetime\ $\tau (\epsilon )$ one
obtains a finite $\rho _{D}^{xy}$. No experiment has reported so far a
finite Hall drag, at least in the situation were one can guarantee that
there is no tunneling between layers. Patel {\it et al.}\cite{patel}
measured frictional drag in the presence of a field\ for modulation doped
GaAs/AlGaAs double quantum wells separated by a barrier of $100$\AA$\;$ for
which tunneling is significant. They found oscillatory behavior as a
function of magnetic field in both the longitudinal and transverse drag
resistivities. However. since for this interlayer separation tunneling is
significant\cite{brown}, our analysis does not apply to this case.

{\it \bigskip }

\subsection{Magneto Coulomb drag}

\bigskip The physics of electrons in two dimensions under quantum conditions
of temperature and magnetic field has been the subject of intense research
in recent years, and the reader is referred to some of the very good   
reviews on the progress in understanding of the integer and fractional 
Quantum Hall Effects\cite{pinczuk,girvin}. In this section we review
the research on current drag in the presence of magnetic fields. 

In the presence of a magnetic field, the polarization function $\chi
(q,\omega )$ assumes different forms depending on the number of filled
Landau levels and on the strength of disorder.
For example, for non--interacting electrons, if  the density and magnetic fields are 
such that  the outmost occupied Landau level is completely 
filled,  the lowest energy for a particle--hole excitation is $\hbar \omega_c$,
with $\omega_c=eB/mc$ the cyclotron frequency. Therefore ${\rm Im} chi$ will consist
of a series of delta functions separated by $\omega_c=eB/mc$. If the outmost
occupied Landau level is partially filled there will be excitations or zero
frequency that correspond to transitions between orbits in the same Landau level.
The presence of disorder smears this delta functions.
One therefore expects that
the transresistance will be sensitive to Landau level quantization for
fields large enough that the cyclotron frequency 
is much larger than the disorder induced lifetime of the orbits. This
problem was studied by B\o nsager {\it et al.}\cite{bonsager1,bonsager2,geldart} and
 Wu\cite{wu} and Qin\cite{qin}, who treated the individual layers as
non--interacting electrons in the presence of disorder consisting of short
range impurities. In the presence of disorder the one--particle density of
states $g(\epsilon )$ is broadened, and B\o nsager {\it et al. }chose to
substitute the comb of delta functions centered at $\epsilon _{n}=(n+\frac{1%
}{2})\hbar \omega _{c}${\it \ }for a Gaussian density of states as derived
by Gerhardts \cite{gerhardts} 
\begin{equation}
g(\epsilon )=\frac{\sqrt{2/\pi }}{2\pi \ell _{B}^{2}\Gamma _{0}}\sum_{m}\exp %
\left[ -2\left( \frac{\epsilon -\epsilon _{m}}{\Gamma _{0}}\right) ^{2}%
\right] ,  \label{gaussian}
\end{equation}
with $\ell _{B}=\sqrt{\hbar /eB\text{ }}$ the magnetic length, and $\Gamma
_{0}=(2/\pi )\hbar \omega _{c}(\hbar /\tau )$, $\tau $ being the transport
lifetime in the absence of magnetic field. The equations for the
transresistance were solved numerically and revealed a very interesting
``twin peak'' structure as a function of filling fraction: as the filling
factor is changed from an odd value (where the highest Landau level is half
filled in the spin--unpolarized situation) towards the even value, the
transresistance goes through a maximum before it is suppressed. The authors
explain this non-monotonic effect as being the result of the competition
between phase space for scattering and the strength of the effective
interaction. When the Landau levels are completely filled, the chemical
potential is in a gap, there is no dissipation, and the transresistance
vanishes. This corresponds to the plateau regions of the Quantum Hall
Effect. \ Also, \ the screening is strongly depressed in the region where
the density of sates is small. In the transition region there are
excitations of zero energy, and since density of states increases as one
approaches the center of the band (half filled Landau levels) one gets an
increase in $\chi (q,\omega )$ contributing to an increase in $\rho _{D}$.
On the other hand, an increase in $\chi (q,\omega )$ means an enhancement of
screening, or equivalently a decrease in the effective interaction. The
prediction is therefore that $\rho _{D}$ should roughly fulfill the relation 
\begin{equation}
\rho _{D}=g_{1}g_{2}|W_{12}|^{2},  \label{rough}
\end{equation}
with $g_{i}$ the density of states at the Fermi energy for layer $i$, and $%
W_{12}$ the effective inter--layer interaction. The striking twin--peak
structure was observed in the experiments by Rubel {\it et al.}\cite{rubel}
with good quantitative agreement with the theory. The twin--peak structure
is observed up to a filling fraction $\nu =15$. There is a marked increase
in the value of\ the transresistance from $\rho _{D}=8{\rm m}\Omega  
$ at zero field and $T=3.1K$ to values of the order of \ $1-2\Omega 
$, as expected from the theory (see Figure 4). The calculation does not include effect of
localization and one should keep in mind that the $g_{i\text{ }}$of Equation
(\ref{rough}) refers to a density of {\em extended }states. The enhancement
of $\rho _{D}$ in the critical region was studied by Shimshoni and Sondhi
\cite{sondhi}. They show that anomalously slow relaxation
of density fluctuations at criticality yields a power--law $\rho_D\sim
T^{2-\eta}$, where $\eta$ is the anomalous diffusion exponent.

The temperature dependence of $\rho _{D}$ \ also shows different behavior
with or without an applied magnetic field. In the discussion of Section \ref
{disorder} we found a low temperature dependence $\rho _{D}\sim T^{2}\ln T$
in the diffusive regime. The range of temperatures for which this dependence
applies corresponds to $k_{B}T\lesssim \hbar /\tau $, with $\tau $ a
scattering time. This defines a diffusive temperature that for high mobility
samples is of the order of 50 mK. Moreover, Zhen and MacDonald\cite{zheng}
estimate--after including the different distance dependences in the
ballistic regime and the diffusive regime--a crossover temperature of $%
10^{-100}K$ ! At low temperatures \ for $B\neq 0$ one expects
the same $T^{2}\ln T$ \ behavior since the motion is also diffusive, but the
temperature scale where the effect sets in can be higher and experimentally
accessible. The reason is that in the presence of a field one can think of
diffusion as hopping between adjacent orbits of radius $\ell _{B}\ll
v_{F}\tau $. This implies that the diffusive form of the polarizability of
Equation (\ref{diffusive}) is now valid for $q\lesssim 1/\ell _{B}$ . The
numerical solutions of B\o onsager {\it et al.}\cite{bonsager1} indicate
that, for a choice of parameters corresponding to identical layers with a
density $n=3\times 10^{15}m^{-2\text{ }}$and an interlayer distance $d=800$%
\AA , the diffusive behavior sets in at $T=0.4$K.

Another very interesting aspect of the temperature dependence at finite
magnetic fields predicted by B\o onsager {\it et al.}\cite{bonsager1} and
observed in the experiments by Rubel {\it et al.}\cite{rubel} is the dynamic
screening originating a maximum in $\rho _{D}/T^{2}$, the maximum being at $%
T_{F}$. The theoretical reason is the same leading to plasmon enhancement
discussed in Section \ref{plasmons}.

Wu\ {\it et al.}\cite{wu}also studied the interplay between Landau
quantization and transresistivity by solving numerically Equation (\ref
{central}) in the presence of a magnetic field. They find pronounced
oscillations as a function of field but not the ``twin--peak'' structure.
Also, experiments by Hill{\it \ et al.}\cite{hill} show oscillations in $%
\rho _{D}$ without the ``twin--peak'' structure. The oscillations are
pronounced at low temperatures $T\lesssim 4K.$ For high temperatures the
Landau quantization effects are washed out and $\rho _{D}$ has a dependence 
\[
\rho _{D}\propto TB^{2}, 
\]
which has not been addressed theoretically yet.

\subsection{The case of $\protect\nu =1/2$}

\bigskip

The case of Landau levels of filling fraction $\nu =1/2$ (half filled) or
even denominators  is special\cite{pinczuk}. The quantized Hall effect does
not occur in this case, and the low energy physics\ is that of a Fermi
liquid, behaving in many ways as electrons in \ zero magnetic
 field\cite{leeread}. The theoretical approach that has proved
most useful for understanding this state is the fermion Chern--Simons theory
\cite{fradkin}, which is based in turn in the composite fermion theory
developed by J. K. Jain\cite{jain}. 

For the fractional Hall effect, Orgad and Levit\cite{orgad} and Duan\cite
{duan/e} studied the Coulomb drag for edge excitations using a Chern--Simons
theory. The case of interlayer friction was considered for $\nu =1/2$ by
Sakhi\cite{sakhi} , Ussishkin and Stern\cite{ussishkin} and Kim and Millis 
\cite{millis}. The dominant low temperature for $\rho _{D}$ is found to be $%
T^{4/3}$. This temperature dependence results from the slow diffusion of the
density modes at filling fraction $1/2$. In the
composite fermion picture, at $\nu =1/2$ the density response at small
frequencies and small wave vectors is of the form\cite{leeread} 
\begin{equation}
\chi (q,\omega )\approx \frac{q^{3}}{q^{3}/\chi _{0}-8\pi i\hbar \omega k_{F}%
},
\end{equation}
with $\chi _{0}=dn/d\mu $ the electronic compressibility. The form of is
similar to that of the diffusive regime at $B=0$ of Equation (\ref{diffusive}%
) with an effective diffusion constant that vanishes linearly in $q$. This
means that the long wavelength density fluctuations relax very slowly
leading to an increase in the transresistance. To obtain the low temperature
dependence we proceed in the same way as for the diffusive case, except that
now, for small $q,$ $%
\mathop{\rm Im}%
\chi \sim 1/q^{3}$, with the divergence now being cutoff at $q\approx
2\left( \chi _{0}\hbar k_{F}\right) ^{1/3}\omega ^{1/3}\equiv k_{F}.(\omega
/\omega _{0})^{1/3}.$ \ The low temperature behavior is given by 
\begin{equation}
\rho _{D}\sim \frac{e^{2}}{(k_{{\rm TF}}d)^{2}}\beta \int_{0}^{\infty
}d\omega \frac{\omega ^{2}}{e^{\beta \hbar \omega }+e^{-\beta \hbar \omega
}-2}\int_{k_{F}(\omega /\omega _{0})^{1/3}}dq\frac{1}{q^{3}}\approx \frac{%
e^{2}}{(k_{{\rm TF}}d)^{2}}T^{4/3}  \label{4/3}
\end{equation}

\bigskip

The case of $\nu =1/2$ poses some fundamental questions that are unresolved
at the moment of writing this review. The recent experiments by Lilly {\it %
et al.}\cite{lilly}{\it \ }for modulation doped GaAs/Al$_{x}$Ga$_{1-x}$As
double quantum wells separated by 200\AA , the drag resistance $\rho _{D}$
has a qualitatively similar behavior as the longitudinal resistance of a
single isolated layer $\rho _{xx}$ when the magnetic field is varied: where $%
\rho _{xx}$ is maximum due to the quantum Hall effect (for example at $\nu
=1 $ and $2/3$), so is $\rho _{D}.$ As pointed out in Ref. \cite{lilly} this
is not surprising since both resistances are controlled by the density of
states available for scattering. There is, however, a notable difference in
their respective temperature dependences. Whereas $\rho _{xx}$ increases by
only $6\%$ as the temperature is lowered from $T=4K$ to $0.2K,$the drag
resistance decreases by a factor of $40$ in this temperature range. The
numerical value of the drag resistance at $B=11{\rm T},$ ( the field
corresponding to $\nu =1/2)$ and at $4K$ is about $2000$ times larger than $%
\rho _{D}$ at zero field. In addition, the temperature dependence of $\rho
_{D}$ differs qualitatively at $\nu =1/2$ from the corresponding resistance
at zero field:\ the experiment does not show evidence of a ``phonon peak''
in $\rho _{D}/T^{2},$ suggesting that phonons are relatively unimportant as
a scattering mechanism in \ the $\nu =1/2$ case.

The most intriguing part of this story is the evidence of a finite value of $%
\rho _{D}\approx 5\Omega $ when extrapolated to zero temperature
(see Figure 5).
This is clearly in conflict with the above discussion leading to the
temperature dependence of Eq. (\ref{4/3}) that predicts a vanishing
resistance at zero temperature. Since the scattering giving rise to $\rho
_{D} $ is inelastic, and the common view is that all inelastic processes
cease to be effective at zero temperature, this experiment is inviting us to
reanalyze the mechanisms of interlayer dissipation at zero temperature. This
interesting experiment places current drag in the same arena as some recent
efforts in understanding the issues of dephasing\cite{mohanty,vavilov} and
resistance at zero temperature\cite{liboff} due to ``inelastic'' mechanisms.
Some very recent papers address  the $\nu =1/2$ from different angles:
Ussishkin and Stern\cite{stern_2} attribute the anomaly to pairing fluctuations
whereas Yang\cite{yang} suggests that the low temperature behaviour of
the drag resistance is due to the interlayer distance being close to the
critical value where the two layers form a collective incompresible state.

\bigskip

\section{Non dissipative drag}

\label{nondisip}

\bigskip The possibility of a drag effect at zero temperature was considered
by Rojo and Mahan\cite{rojo1}, who considered two coupled \ meosocopic\cite
{mesoscopics} rings that can individually sustain persistent currents. The
mechanism giving rise to drag in a non--dissipative system is also based on
the inter--ring or inter--layer Coulomb interaction, the difference with the
dissipative case being the coupling between real or virtual interactions.
One geometry in which this effect comes to life is two collinear rings of
perimeter $L$, with a Bohm--Aharonov $\Phi _{1}$ flux threading {\em only one%
} of the rings (which we will call ring one). This is of course a difficult
geometry to attain experimentally, but has the advantage of making the
analysis more transparent. Two coplanar rings also show the same effect\cite
{rojo1}. If the rings are uncoupled in the sense that the Coulomb
interaction is zero between electrons in different rings, and the electrons
are non--interacting within the rings, a persistent current $%
J_{0}=-cdE/d\Phi _{1}=ev_{F}/L$ will circulate in ring one\cite{buttiker}.
If the Coulomb interaction between rings is turned on, the Coulomb
interaction induces coherent charge fluctuations between the rings, and the
net effect is that ring two acquires a finite persistent current. The
magnitude of the \ persistent drag current $J_{D}$ can be computed by
treating the modification of the ground state energy in second order
perturbation theory $\Delta E_{0}^{(2)}$, and evaluating 
\begin{equation}
J_{D}=-e\left. \frac{d\Delta E_{0}^{(2)}}{d\Phi _{2}}\right| _{\Phi _{2}=0},
\label{persist0}
\end{equation}
with $\Phi _{2}$  an auxiliary flux treading ring two that we remove after
computing the above derivative.
In other words, the persistent drag current 
is equivalent to a induced diamagnetic current that is finite even when 
the flux is zero in that system. (Note that a diamagnetic current is
in general given
by the differential change in energy with respect to a change in magnetic
field.)
The correction to the energy resembles the
van--der Waals interaction, and its relevance to systems that can
individually break time reversal symmetry was studied in Ref.\cite
{rojo2,rojo3}. The second order correction is given by 
\begin{equation}
\Delta E_{0}^{(2)}=-\sum_{q}|U_{q}|^{2}\int_{0}^{\infty }d\omega
\int_{0}^{\infty }d\omega ^{\prime }\frac{S_{1}(q,\omega )S_{2}(-q,\omega
^{\prime })}{\omega +\omega ^{\prime }},
\end{equation}
with $S_{i}(q,\omega )$ the dynamical structure factor of ring $i$ (see Eq.%
\ref{perturb}). For a mesoscopic ring with an applied Bohm--Ahronov flux one
has to retain the discreteness of the spectrum: 
\begin{equation}
S_{i}(q,\omega )=S_{i}^{(0)}\left( q,\omega -\frac{\hbar ^{2}}{2m}q\frac{%
\phi _{i}}{2\pi L}\right) ,
\end{equation}
with $S_{i}^{(0)}(q,\omega )$ the structure factor at zero flux, $\phi
_{i}=\Phi _{i}/\phi _{0},$ and $\phi _{0}=hc/e$ the flux quantum. Note that,
for a mesoscopic system, due to the presence of the flux, $%
S_{i}^{(0)}(q,\omega )\neq S_{i}^{(0)}(-q,\omega ).$ For the inter--ring
interaction we take the unscreened Coulomb interaction (screening is not
effective in one--dimension), so $U_{q}=K_{0}(qd)$, with $d$ the distance
between rings and $K_{0}(x)$ the modified zeroth--order Bessel Function. In
the limit of the interparticle distance much smaller than the distance
between rings ($k_{F}d\gg 1$), we obtain 
\begin{equation}
J_{D}=J_{0}\frac{1}{(k_{F}a_{0})^{2}}\frac{1}{(k_{F}d)^{2}},  \label{jdrag}
\end{equation}
with $a_{0}$. For dimensions corresponding to the experiments measuring
persistent currents\cite{levy} $J_{D}\simeq 10^{-4}J_{0}$. The drag current
is itself mesoscopic, and therefore vanishes in the limit of infinite
length. For a ring with a single channel carrying the current, an extension
of the argument presented by Vignale\cite{vignale1} for a bound in the value
of the persistent currents gives $J_{D}<J_{0}$ in general. The
non--dissipative drag for two concentric rings was studied by Shahbazyan and
Ulloa\cite{ulloa} using a Luttinger liquid formulation of mesoscopic systems 
\cite{loss}. They found that the inter--ring interaction modifies the period
of the Aharonov--Bohm osillations. Related work on the effect of the
interactions on the flux dependece was reported by Canali {\it et al}.\cite
{canali}.

One basic difference between the dissipative drag in semiconductor
systems  and the non--dissipative drag in
mesoscopic systems appears if one opens ring two so that no current can
circulate, and computes and induced charge modulation, which will play the
role of a drag voltage. The induced ``voltage'' is zero in this case. This
can be seen by starting with a setup that in the absence of flux in system
one is ``parity even''. By this we mean that the charge distribution in wire
two is symmetric around the center. Let us call ${\cal T}$ and ${\cal P}$
the time reversal and parity operators respectively. We want to compute the
induced dipole moment in ring two $x_{2}=\langle \Psi _{0}|\widehat{x}%
_{2}|\Psi _{0}\rangle $. The coordinate operator satisfies ${\cal PT}%
\widehat{x}_{2}({\cal PT})^{-1}=-\widehat{x}_{2}$ while the wave function is
invariant under ${\cal PT}$, which implies $x=0$. The charge distribution in
ring one remains uniform, and there is no induced voltage.

A natural candidate for the study of non--dissipative drag other than
mesoscopic systems are superconductors, which can sustain macroscopic
persistent currents. The extension to this case was done by Duan and Yip\cite
{duan}. For wires the corrections to the zero point energy of the charge
fluctuations (plasmon modes) was computed when two superconducting wires
individually carry a supercurrent. \ In analogy with the discussion of
Section \ref{interlayer} the dispersion of the coupled modes is given by the
determinantal equation 
\begin{equation}
\left| 
\begin{array}{cc}
(\omega -qv_{1})^{2}-s^{2}q^{2} & -X \\ 
-X & (\omega -qv_{1})^{2}-s^{2}q^{2}
\end{array}
\right| =0,  \label{duan1}
\end{equation}
with $s$ the plasmon velocity of a single wire, $X=4\pi
n_{0}e^{2}q^{2}K_{0}(qd)/m$ comes from the inter--wire interaction, and $%
v_{1,2\text{ }}$are the superconducting velocities of each wire. For the
case of wires the two couple modes of frequencies $\omega _{\pm }(q)$ are
linear in $q.$ The zero point energy $E_{0}=\sum_{q}\frac{1}{2}\left[ \hbar
\omega _{+}(q)+\hbar \omega _{-}(q)\right] $ depends now on the relative
superfluid velocities. The superfluid velocity is $\hbar (\nabla \Psi -2e{\bf %
A}/c)/2m$ with $\Psi $ the order parameter. The supercurrent in wire two $%
I_{2}$ is computed in an analogous way as Eq. (\ref{persist0}) by taking the
derivative of the free energy with respect to the vector potential in wire
2, and 
\[
I_{2}=\frac{e}{m}(\rho _{22}v_{2}+\rho _{21}v_{1}),
\]
with $\rho _{12}=\hbar n_{0}^{2}e^{4}/16\pi m^{2}s^{5}d^{2}$ representing
the drag term. If one starts with a situation in which there is no current
in either wire and slowly increases the current in wire one, the prediction
is that a current will start to flow in wire two of magnitude $e\rho
_{21}v_{1}/m$. Superconductivity is essential to this effect, so that the
wire can be trapped in a metastable state. Extensions of the above arguments
to superfluid Bose systems were presented by Shevchenko and Terent'ev\cite
{shevchenko} and Tanatar and Das\cite{tanatar}. The case of the
transresistance of an excitonic condensate with electrons in one layer and
holes in the other layer was studied by Vignale and MacDonald\cite{vignale2}%
, who found a discontinuous jump in $\rho _{D}$ at the condensation
temperature.

Two groups attempted to measure the non--dissipative drag. Giordano and
Monnier\cite{giordano} measured the drag between a superconductor (Al)\ and
a normal metal (Sb), and found a non--reciprocal drag effect that was finite
only in the transition region, that is, at temperatures close to the
critical temperature of the Al layer. Similar results were reported by Huang 
{\it et al}.\cite{huang} with the superconducting system being AlO$_{x}$ and
Au for the normal metal. An interpretation of the non--reciprocity effect in
terms of inductive coupling of the spontaneously generated vortices in the
superconductor and the normal metal was proposed in Ref.\cite{shimshoni2}.
It seems clear that these experiments do not provide evidence of the
supercurrent drag.

\bigskip\ \ 

\section{Summary}

We have reviewed the recent theoretical and experimental efforts in the growing field of
current drag. 
For electronic  systems separated far enough so that there is no tunneling
between them, 
we found that through studies of the transresistance $\rho_D$ one 
can extract information not only about the direct Coulomb interaction between
electrons in different systems,  but also about the collective modes of the coupled systems, 
and about phonons (virtual and real) that can propagate through the barrier separating
the systems in question. The magnetic field $B$ has a non-trivial effect on the drag
resistance. The value of $\rho_D $ is larger af finite fields than its value at zero
field, and shows a large enhancement in the transition region between
quantum Hall plateaus. This effect is understood qualitatively in terms of
 an interplay between changes 
in phase space available for scattering and variations in the effective interaction with $B$. 
Some theoretical considerations, like those in connection with the non--dissipative
drag, and the ones for strongly disordered systems, are still waiting  experimental 
verification. The recent experiment on the 
$\nu=1/2$ case pose a fundamental question connected to the general theory of transport:
 is it possible to have an intrinsic resistance at zero temperature due to the electron--electron
 interaction? The magnitude of the challenge seems as big as that of the progress made so 
 far in the field. 

\bigskip\
We acknowledge interactions with   W. Dietsche, J. P. Eisenstein, A. J. Leggett,
A. H. MacDonald, 
 G. D. Mahan, C. Proetto, E. Shimshoni, S. Ulloa, P. Vasilopoulos and G. Vignale. 

\bigskip 

\appendix
\section{}
In this section we follow the derivation of Zheng and MacDonald\cite{zheng}
of the interplanar resistance using the memory function formalism. The
formalism of Mori\cite{mori} uses the projector technique to write the
response function in terms of a memory function. A simple case is a
classical particle subject to stochastic forces, for which one is interested
in the velocity--velocity correlation function $\phi (t)=\langle
v(t)v(0)\rangle $. The Laplace transform is 
\begin{equation}
\phi (s)={\frac{\langle v^{2}(0)\rangle }{s+K(s)}},
\end{equation}
with $K(s)$ the Laplace transform of the memory function 
\begin{equation}
K(t)={\frac{1}{mk_{B}T}}\langle F(t)F(0)\rangle .
\end{equation}
Here $F(t)$ is the force acting on the particle, and $m\langle
v^{2}(0)\rangle =k_{B}T$. The friction coefficient is then given by a
force--force correlation function.

For the quantum mechanical case, the calculation of an inter--layer
resistance is then performed by identifying the relevant memory function.
The starting point is Kubo's formula for the conductivity: 
\begin{equation}
\sigma _{ij}(\omega )={\frac{\beta }{A}}\int_{0}^{\infty }dte^{i\omega
t}\langle J_{j}|J_{i}(t)\rangle  \label{cucu}
\end{equation}
where $\beta =1/k_{B}T$, $A$ is the area of each of the two--dimensional
systems considered, $i$ and $j$ are the layer indices, and $J$ is the zero
wave vector component of the total current operator. The inner product
appearing in (\ref{cucu}) is 
\begin{equation}
\langle \hat{J}_{j}|\hat{J}_{i}(t)\rangle ={\frac{1}{\beta }}\int_{0}^{\beta
}d\lambda \;{\rm Tr}\left[ \rho _{0}\exp (\lambda H)\hat{J}_{j}\exp
(-\lambda H)\hat{J}_{i}(t)\right] ,
\end{equation}
with $\hat{J}_{i}(t)=e^{{\frac{i}{\hbar }}Ht}\hat{J}_{i}e^{-{\frac{i}{\hbar }%
}Ht}=e^{i{\cal L}t}\hat{J}_{i}$, and ${\cal L}$ is the Liouville
superoperator defined through its action on an operator $\hat{O}$ as ${\cal L%
}\hat{O}={\frac{1}{\hbar }}[H,\hat{O}]$. The next step is to use Mori's
projector method to write the inverse matrix $\sigma _{i,j}^{-1}(\omega )$
in terms of a memory function, and hence obtain transresistance. Zhen and
MacDonald define a superoperator ${\cal P}$ that projects onto the current: 
\begin{equation}
{\cal P}={\frac{|J_{1}\rangle \langle J_{1}|}{\langle J_{1}|J_{1}\rangle }}+{%
\frac{|J_{2}\rangle \langle J_{2}|}{\langle J_{2}|J_{2}\rangle }}\equiv 1-%
{\cal Q},
\end{equation}
with $\langle J_{i}|J_{j}\rangle =\delta _{ij}{\frac{A}{\beta }}\chi _{i}$,
and $\chi _{i}=n_{i}e^{2}/m$.

With this definitions we have (for ${\rm Im}\omega >0$), 
\begin{eqnarray}
\sigma _{ij}(\omega ) &=&{\frac{\beta }{A}}\int_{0}^{\infty }dte^{i\omega
t}\langle J_{i}|e^{-i{\cal L}t}|J_{j}\rangle ={\frac{\beta }{A}}\langle
J_{i}|{\frac{i}{\omega -{\cal L}}}|J_{j}\rangle \\
&=&{\frac{\beta }{A}}\langle J_{i}|{\frac{i}{\omega -{\cal QL}-{\cal PL}}}%
|J_{j}\rangle \\
&=&{\frac{\beta }{A}}\langle J_{i}|{\frac{i}{\omega -{\cal QL}}}%
|J_{j}\rangle +{\frac{\beta }{A}}\langle J_{i}|{\frac{i}{\omega -{\cal L}}}%
{\cal PL}{\frac{1}{\omega -{\cal QL}}}|J_{j}\rangle \\
&=&\delta _{ij}{\frac{i\chi _{i}}{\omega }}+\sum_{k=1}^{2}{\frac{\sigma
_{ik}(\omega )}{\omega }}{\frac{\beta }{A}}{\frac{1}{\chi _{k}}}\langle
J_{k}|{\cal L}{\frac{i}{\omega -{\cal QL}}}{\cal L}|J_{j}\rangle ,
\end{eqnarray}
where we have used ${\cal P}{\dot{J}_{i}}=i{\cal P}{\cal L}{\ J_{i}}=0$,
which results, in the present case, from the fact that $[J_{i},J_{j}]=0$,
meaning that the derived expressions are also valid in the presence of a
magnetic field that breaks time reversal invariance. We are interested in
the real part of $\sigma ^{-1}$, which gives the resistance. From the above
equations we have 
\begin{eqnarray}
\rho _{12}(\omega ) &=&{\frac{\beta }{A}}{\frac{1}{\chi _{1}\chi _{2}}}{\rm %
Re}\langle \dot{J}_{1}|{\frac{i}{\omega -{\cal QL}}}|\dot{J}_{2}\rangle \\
&=&{\frac{\beta }{A}}{\frac{1}{n_{1}n_{2}e^{2}}}\int_{0}^{\infty
}dte^{i\omega t}\langle F_{1}|e^{-i{\cal QL}t}|F_{2}\rangle ,  \label{ff}
\end{eqnarray}
where the force operator is $F_{i}=-m/e\dot{J}_{i}$. The contribution to the
force due to the inter--layer interaction potential $U_{{\bf q}}$ is 
\begin{equation}
{\bf F}_{1}=-{\bf F}_{2}={\frac{1}{A}}{\frac{i}{\hbar }}\sum {\bf q}U_{{\bf q%
}}\rho _{{\bf q}}^{(1)}\rho _{-{\bf q}}^{(2)}.  \label{force}
\end{equation}

The first approximation is to replace $e^{-i{\cal Q L}}$ by $e^{-i{\cal L}}$%
. The leading order in $U_{{\bf q}}$ corresponds to the correlation function
in (\ref{ff}) evaluated in the uncoupled case. Using a representation in
terms of exact eigenstates, one finds (for the static limit)

\begin{equation}
\rho _{12}={\frac{\hbar ^{2}\beta }{\pi n_{1}n_{2}e^{2}A}}\int {\frac{d{\bf q%
}}{(2\pi )^{2}}}q^{2}U_{{\bf q}}\int_{-\infty }^{\infty }d\omega {\frac{{\rm %
Im}\chi _{1}(q,\omega ){\rm Im}\chi _{2}(q,\omega )}{\sinh ^{2}(\beta \hbar
\omega /2)}},  \label{central0}
\end{equation}
with $\chi _{i}(q,\omega )$ the density--density response function\cite
{nozieres} of layer $i$: 
\begin{eqnarray}
{\rm Im}\chi _{i}(q,\omega ) &=&{\frac{\pi }{A}}(1-e^{\hbar \omega \beta }){%
\frac{1}{Z_{i}}}\sum_{m,n}e^{-\beta E_{m}}|\langle n|\rho _{i}(q)|m\rangle
|^{2}\delta \lbrack \omega -(E_{m}-E_{n})/\hbar ]  \label{perturb} \\
&\equiv &(1-e^{\hbar \omega \beta })S(q,\omega ),
\end{eqnarray}
and $Z_{i}$ the partition function of the isolated layer.

Here we show that $\rho_{12}$ vanishes as $T^2$. For simplicity we treat the
layers as identical. The approximation used in the literature\cite
{gramila0,mahan0} is to replace $\chi_i(q,\omega)$ by its non--interacting,
zero temperature value\cite{stern}:

\begin{equation}
{\rm Im}\chi (q,\omega )={\frac{m}{\pi q\hbar ^{2}}}\left[ \Theta
(k_{F}-|x_{-}|)\sqrt{k_{F}^{2}-x_{-}^{2}}-\Theta (k_{F}-|x_{+}|)\sqrt{%
k_{F}^{2}-x_{+}^{2}}\right] ,
\end{equation}
with $x_{\pm }=m\omega /\hbar q\pm q/2$. At low temperatures, the factor $%
1/\sinh ^{2}(\beta \hbar \omega /2)$ in (\ref{central}) ensures that the
values of $\omega $ are small. Then we can separate the density response
functions as 
\begin{equation}
{\rm Im}\chi (q,\omega )={\frac{2m^{2}\Theta (2k_{F}-q)}{\pi \hbar ^{3}q%
\sqrt{(2k_{F})^{2}-q^{2}}}}\omega \equiv F(q)\omega .
\end{equation}
The integral over $\omega $ is 
\begin{equation}
\beta \int_{-\infty }^{\infty }d\omega {\frac{\left[ {\rm Im}\chi (q,\omega )%
\right] ^{2}}{\sinh ^{2}(\beta \hbar \omega /2)}}=(k_{B}T)^{2}{\frac{4\pi
^{2}}{3\hbar ^{3}}}F(q)^{2}.
\end{equation}
The remaining integral is 
\begin{equation}
I=\int dq\;q^{3}U_{b}^{2}(q)F(q)^{2}.
\end{equation}
Here, one can replace the bare interlayer interaction with 
\begin{equation}
U_{b}(q)\rightarrow U_{{\bf q}}^{(SC)}={\frac{2\pi e^{2}}{q}}{\frac{e^{-qd}}{%
\epsilon _{\infty }\epsilon (q)}},
\end{equation}
where $\epsilon (q)$ is the effective dielectric function for the two
parallel conducting planes, and $d$ is the distance between planes. In the
random--phase approximation, 
\begin{equation}
\epsilon (q)=1-2U_{b}(q)\chi (q,0)+U_{b}^{2}(q)\chi ^{2}(q,0)(1-e^{-2qd}).
\end{equation}

In the long wavelength limit $U_{b}(q)\chi (q,0)=-q_{{\rm TF}}/q$, with $q_{%
{\rm TF}}=2/a_{0}^{\ast }$ the screening wave vector in two dimensions. The
last term in $\epsilon (q)$ dominates whenever $q_{{\rm TF}}\gg 1$. In this
limit, the integral $I$ can be approximated as 
\begin{eqnarray}
I &\simeq &{\frac{4e^{4}m^{4}}{\epsilon _{\infty }^{2}\hbar ^{8}k_{F}^{2}q_{%
{\rm TF}}^{4}}}\int_{0}^{\infty }q^{3}dq{\frac{e^{-2qd}}{(1-e^{-2qd})^{2}}}
\\
&=&{\frac{3m^{4}\zeta (3)}{2\epsilon _{\infty }^{2}\hbar ^{8}k_{F}^{2}q_{%
{\rm TF}}^{4}d^{4}}}.
\end{eqnarray}
Collecting this integrals we obtain 
\begin{equation}
\rho _{12}={\frac{m}{ne^{2}}}{\frac{\pi \zeta (3)(k_{B}T)^{2}}{16\hbar
E_{F}(q_{{\rm TF}}d)^{2}(k_{F}d)^{2}}}
\end{equation}

\begin{figure}
\label{fig1}
\caption{ 
Schematic representation of the memory function expression for the transresistance to
lowest order in the interlayer interaction [Eq.(\ref{central})] [reproduced from Ref. (\cite{zheng})]}
\end{figure}

\begin{figure}
\label{fig2}
\caption{The scattering rate due to the Coulomb scattering and virtual phonons $\tau_D^{-1}/T^2$ as
a function of temperature for different separations. Note that $\rho_D \propto \tau_D^{-1}$. The 
solid circles are the experimental results of Ref. (\cite{gramila/prl}) and the solid lines are
the theoretical results from Ref. (\cite{tso0}). Inset: Contribution to $\rho_D \propto \tau_D^{-1}$
due to exchange of virtual phonons as a function of temperature. [Reproduced from Ref. (\cite{tso0})].}
\end{figure}

\begin{figure}
\label{fig3}
\caption{ 
The scaled transresistivity $\rho_t/T^{-2}$ ($\rho_t \equiv \rho_D$) versus the reduced temperature
for different densities (the densities in both layers are the same). The dashed (solid) lines are
the RPA (Hubbard) calculations of Flensberg and Hu\cite{flensberg2}, and the circles 
are the experimental results of Ref. (\cite{hill2}). [Reproduced from Ref. (\cite{hill2}).]
}
\end{figure}

\begin{figure}
\label{fig4}
\caption{ The transresistance $R_T$ ($\rho_D$) as a function of magnetic field $B$ for a coupled 
electron gas with a separation barrier of 30 nm. shown for different temperatures (plotted with
offset for clarity). The electron density in both layers is $n=3.2 \times 10^{11}$cm$^{-2}$.  The 
longitudinal resistance is also shown. 
[Reproduced from Ref. (\cite{rubel}).]
}
\end{figure}
\begin{figure}
\label{fig5}
\caption{ Measured temperature dependence of $\rho_D$ at $\nu=1/2$ (solid line). The broken lines
are calculations from Ref. \cite{sakhi,ussishkin} of $\rho_D$ assuming two different values of the
composite fermions mass (dotted, $m^*=12 m_b$, dashed, $m^*=4m_b$, where $m_b$ is the GaAs band mass).
[Reproduced from Ref. (\cite{lilly}).]
}
\end{figure}


\begin{references}

\bibitem{gramila/prl}  T. J. Gramila, J. P. Eisenstein, A. H. MacDonald, L.
N. Pfeiffer and K. W. West, Phys. Rev. Lett. {\bf 66}, 1216 (1991).


\bibitem{solomon1}  P. M. Solomon, P.J. Price, D. J. Franck and D. C. La
Tulipe, Phys. Rev. Lett. {\bf 63 }2508 (1989).

\bibitem{gramila0}  T. J. Gramila, J. P. Eisenstein, A. H. MacDonald, L. N.
Pfeiffer and K. W. West, Phys. Rev. B {\bf 47}; Physica B {\bf 197}, 442
(1994).

\bibitem{pogre}  M. B. Pogrebrinski\u{\i}, Fiz. Tekh, Poluprovodn. {\bf 11},
637 (1977) [Sov. Phys. Semicond. {\bf 11}, 372 (1977)]


\bibitem{price0}  P.J. Price, Physica {\bf 117}/{\bf 118 B}$+${\bf C}, 750
(19
83).

\bibitem{santoro}  G. E. Santoro and G. F. Giuliani, Phys. Rev. B {\bf 37},
937 (1988).

\bibitem{nozieres}  D. Pines and P. Nozi\`{e}res, {\em The theory of Quantum
Liquids}, Vol I. (Addison Wesley, Reading, MA, 1989) Chapter 2.
   

\bibitem{dassarma}  S. D. Sarma and A. Madhukar, Phys Rev. B {\bf 23}, 805
(1981).


\bibitem{ussishkin}  I. Ussishkin and A. Stern Phys. Rev. B {\bf 56}: 4013
(1997).


\bibitem{stern}  F. Stern, Phys. Rev. Lett. {\bf 18}, 546 (1967).


\bibitem{sivan}  U. Sivan, P. M. Solomon and H. Shtrikman, Phys. Rev. Lett. 
{\bf 68}, 1196 (1992).

\bibitem{zheng}  L. Zheng and A. H. MacDonald Phys. Rev. B {\bf 48}, 8203
(1993).


\bibitem{jauho}  A.-P. Jauho and H. Smith, Phys. Rev. B. {\bf 47}, 4420
(1993).

\bibitem{solomon}  P. M. Solomon and B. Laikthman, Superlatt. Microstruct.
10, {\bf 89} (1991).

\bibitem{maslov}  D. L. Maslov, Phys. Rev. B {\bf 45, }1911 (1992).


\bibitem{tso}  H. C. Tso and P. Vasilopoulos, Phys. Rev. B {\bf 45}, 1333
(1992);

\bibitem{hansch}  W. H\"{a}nsch and G. D. Mahan, J. Phys. Chem. Solids {\bf 44%
} 663 (1983).


\bibitem{flensberg0}  K. Flensberg, B. Y.-K. Hu, A.-P \ Jauho and J. M.
Kinaret, Phys Rev. B {\bf 52}, 14761 (1995).


\bibitem{gramila0}  T. J. Gramila, J. P. Eisenstein, A. H. MacDonald, L. N.
Pfeiffer and K. W. West, Phys. Rev. B {\bf 47}; Physica B {\bf 197}, 442
(1994).


\bibitem{hodges}  C. Hodges, H. Smith and J. W. Wilkins, Phys. Rev. B {\bf 4,%
} 302 (1971).


\bibitem{hubner}  K. Hubner and W. Schockley, Phys. Rev. Lett {\bf 4}, 504
(1960).

\bibitem{gurzhi}  R. N. Gurzhi and A. I. Kopeliovich, JETP Lett. {\bf 26},
140 (1977).


\bibitem{price}  P. J. Price, Solid State Commun. {\bf 51}, 607 (1984).


\bibitem{zhang1}  C. Zhang and Y. Takahashi, J. Phys. Cond. Mat. {\bf 5},
5009 (1993).

\bibitem{tso0}  H. C. Tso, P. Vasilopoulos and F. M. Peeters, Phys. Rev.
Lett {\bf 68}, 2516 (1992).

\bibitem{mahan}  G. D. Mahan, {\sl Many--Particle Physics}, 2nd. ed. (Plenum
Press, New York, 1990).



\bibitem{bonsager3}  M. C. B\o nsager, K. Flensberg, B. Y.-K. Hu and A. H.
MacDonald, Phys. Rev. B {\bf 57}, \ 7085 (1998).


\bibitem{mahan}  G. D. Mahan, {\sl Many--Particle Physics}, 2nd. ed. (Plenum
Press, New York, 1990).


\bibitem{price2}  P. J. Price, Ann. Phys.{\bf \ 133}, 217 (1981).


\bibitem{lilly}  M. P. Lilly, J. P. Eisenstein, L. N. Pfeiffer and K. W.
West, Phys. Rev. Lett. {\bf 80} 1714 (1998).

\bibitem{leerama}  P. A. Lee and T. V. Ramakrisnan, Rev. Mod. Phys. 57, 287
(1985).

\bibitem{kamenev}  A. Kamenev and Y. Oreg, Phys. Rev. B {\bf 52}, 7516
(1995).


\bibitem{shimshoni1}  E. Shimshoni, Phys. Rev. B {\bf 56}, 13301 (1997).


\bibitem{wollhardt}  D. Wollhardt and P. W\"{o}lfle, Phys. Rev. B {\bf 22, }%
4666 (1980).

\bibitem{pines} Ref.  \cite{nozieres} pp. 210-215.

\bibitem{flensberg1}  K. Flensberg and B. Y.-K. Hu, Phys. Rev. Lett. {\bf 73}%
, 3572 (1994).

\bibitem{flensberg2}  K. Flensberg and B. Y.-K. Hu, Phys. Rev. B, {\bf 52},
14796 (1995).

\bibitem{tanatar3}  See also B. Tanatar, J. Appl. Phys {\bf 81}, 6214 (1997).

\bibitem{hill2}  N. P. R. Hill, J. T. Nicholls, E. H. Linfield, M. Pepper,
and D. A. Ritchie, G. A. Jones, B. Y.-K. Hu and K. Flensberg, Phys. Rev.
Lett. {\bf 78}, \ 2204 (1997).


\bibitem{swierkowski}  L. \'{S}wierkowski, J. Szyma\'{n}ski and Z. W.
Gortel, Phys. Rev. Lett. {\bf 74}, 3245 (1995).

\bibitem{singwi}  K. S. Singwi, M. P. Tosi, A. H. Land, and A. Sj\"{o}%
lander, Phys. Rev. B {\bf 176}, 589 (1968).

\bibitem{zheng2}  L. Zheng and A. H. MacDonald Phys. Rev. B {\bf 49}, 5522
(1994).

\bibitem{swierkowski2}  J. Szyma\'{n}ski, L. \'{S}wierkowski, and D.
Neilson, Phys. Rev. B {\bf 50}, 11002 (1994).

\bibitem{tso1}  H. C. Tso, P. Vasilopoulos and F. M. Peeters, Phys. Rev.
Lett. {\bf 70}, 2146 (1993).

\bibitem{guven}  G\"{u}ven and B. Tanatar, Solid State Comm, {\bf 104}, 439
(1997).

\bibitem{byk}  B. Y. K. Hu. Phys. Scripta {\bf T69} 170-173 (1997).


\bibitem{patel}  N. K. Patel, E. H. Linfield, K. M. Brown, M. Pepper, D. A.
Ritchie and G. A. C. Jones,
 Semicond. Ssc. Tech.
{\bf 12}, 309 (1997).

\bibitem{brown}  K. M. Brown, E. H. Linfield, D. A. Ritchie, G. A. C. Jones,
M. P. Grimshaw and A. C. Churchill, J. Vac. Sci. Technol. {\bf 12}, 1293
(1994).

\bibitem{pinczuk}  S. Das Sarma and A. Pinczuk eds., {\sl Perspectives in
Quantum Hall Effects}, John Wiley \& Sons, New York (1997).

\bibitem{girvin}  R. E. Prange and S. M. Girvin, {\sl The Quantum Hall Effect%
}, Springer Verlag, New York (1997).

\bibitem{bonsager1}  M. C. B\o nsager, K. Flensberg, B. Y.-K. Hu, and A.-P \
Jauho, Phys. Rev. Lett. {\bf 77 }1366 (1996).

\bibitem{bonsager2}  M. C. B\o nsager, K. Flensberg, B. Y.-K. Hu, and A.-P \
Jauho, Phys. Rev. B {\bf 56 } 10314 (1997).

\bibitem{geldart}  H. C. W. Tso, D. J. W. Geldart and P. Vasilopoulos, Phys.
Rev. B {\bf 57}, 6561 (1988).

\bibitem{wu}  M. W. Wu, H. L. Cui, and N. J. M. Horing, Mod. Phys Lett. B 
{\bf 10}, 279 (1996).

\bibitem{qin}  G. Qin, Solid State Commun {\bf 101} 267 (1997).


\bibitem{gerhardts}  R. B. Gerhardts, Z. Phys. B {\bf 21}, 275 (1975), {\bf %
21}, 285 (1975).


\bibitem{rubel}  H. Rubel , A. Fischer , W. Dietsche, K. von Klitzing and K.
Eberl, Phys. Rev. Lett. {\bf 78} 1763 (1997).


\bibitem{sondhi}  E. Shimshoni, S. L, Sondhi, Phys. Rev. B {\bf 49}, 11484
(1994).

\bibitem{hill}  N. P. R. Hill, J. T. Nicholls, E. H. Linfield, M. Pepper,
and D. A. Ritchie, J. Phys. Condens. Matter {\bf 8}, L557 (1996).


\bibitem{leeread}  B. I. Halperin. P. A. Lee and N. Read, Phys. Rev. B {\bf %
47}, 7312 (1993).


\bibitem{fradkin}  A. Lopez and E. Fradkin, Phys. Rev. B {\bf 44}, 5246
(1991)

\bibitem{jain}  J. K. Jain, Phys. Rev. Lett. {\bf 63}, 199 (1989).


\bibitem{orgad}  D. Orgad and S. Levit, Phys. Rev. B {\bf 53}, 7964 (1996).

\bibitem{duan/e}  J.-M. Duan, Europhys. Lett.{\bf \ 29}, 489 (1995).

\bibitem{sakhi}  S. Sakhi Phys. Rev. B {\bf 56}: 4098 (1997).

\bibitem{millis}  J. B. Kim and A. J. Millis, preprint cond-mat/9611125.

\bibitem{mohanty}  P. Mohanty, E. M. Q. Jariwala and R. A. Webb, Phys. Rev.
Lett. {\bf 78}, 3366 (1997).

\bibitem{vavilov}  M. Vavilov and V. Ambegaokar, preprint cond-mat/9709241.

\bibitem{liboff}  R. L. Liboff and G. K. Schenter, Phys. Rev. B {\bf 54},
16591 (1996).

\bibitem{stern_2} I Ussishkin and A. Stern, Phys. Rev. Lett. {\bf 81}, 3932 (1998).

\bibitem{yang} K. Yang, preperint cond-mat/9806153.

\bibitem{rojo1}  A. G. Rojo and G. D. Mahan, Phys. Rev. Lett. {\bf 68}, 2074
(1992).

\bibitem{mesoscopics}  For recent monograhs in mesoscopic systems see:
Y.Imry, {\em Introduction to mesoscopic systems} (Oxford University Press,
1997); S. Datta, {\em Electronic Transport in Mesoscopic Systems} (Cambridge
University Press, 1995).


\bibitem{buttiker}  M. B\"{u}ttiker, Y. Imry and and R. Landauer, Phys.
Lett. {\bf 96A}, 365 (1983).

\bibitem{rojo2}  A. G. Rojo and A. J. Leggett, Phys. Rev. Lett. {\bf 67},
3614 (1991).

\bibitem{rojo3}  A. G. Rojo and G. S. Canright, Phys. Rev. Lett. {\bf 66},
949 (1991).

\bibitem{levy}  L. P. Levy, G. Dolan, J. Dunsmuir and H. Bouchiat, Phys.
Rev. Lett. {\bf 64}, 2074 (1990).

\bibitem{vignale1}  G. Vignale, Phys. Rev. B {\bf 51} 2612 (1995).

\bibitem{ulloa}  T. V. Shahbazyan and S. E. Ulloa, Phys. Rev. B{\bf \ 55},
13702 (1997).

\bibitem{loss}  D. Loss, Phys. Rev. Lett. {\bf 69}, 343 (1992).

\bibitem{canali}  C. M. Canali, W. Stephan, L. Y. Gorelik, R. I. Shekhter
and M. Jonson, Solid State. Comm. {\bf 104}, 75 (1997)

\bibitem{duan}  J. M. Duan and S. Yip, Phys. Rev. Lett. {\bf 70}, 3647
(1993).

\bibitem{shevchenko}  S. I. Shevchenko and S. V. Terent'ev, Low Temp. Phys. 
{\bf 23}, 817 (1997).

\bibitem{tanatar}  B. Tanatar and A. K. Das, Phys. Rev. B {\bf 54}, 13827
(1996).


\bibitem{vignale2}  G. Vignale and A. H. MacDonald, Phys. Rev. Lett. {\bf 76}%
, 2786 (1996).

\bibitem{giordano}  N. Giordano and J. D. Monnier, Phys. Rev. B {\bf 50},
9363 (1994).

\bibitem{huang}  X. Huang, G. Baz\`{a}n and G. H. Bernsteing, Phys. Rev.
Lett, {\bf 74 }4051 (1995).

\bibitem{shimshoni2}  E. Shimshoni, Phys. Rev. B {\bf 51}, \ 9415 (1995).


\bibitem{mori}  H. Mori, Prog. Thor. Phys. (Kyoto) {\bf 34}, 423 (1965). See
also D. Forster, {\em Hydrodynbamic Fluctuations, Broken Symmetry and
Correlation functions} (Addison Wesley, Reading, MA, 1990) Chapter 5; or B.
J. Berne and G. D. Harp, Adv. Chem. Phys. {\bf XVII}, 63 (1970).


\bibitem{mahan0}  G. D. Mahan, private communicacion.



\end{references}
\end{document}